\documentclass[conference]{IEEEtran}
\IEEEoverridecommandlockouts
\usepackage{graphicx}
\usepackage{algorithm,algorithmic}
\usepackage{subfigure}
\usepackage{amsmath,amsfonts}
\usepackage{algorithmic}
\usepackage{algorithm}
\usepackage{array}
\usepackage[caption=false,font=normalsize,labelfont=sf,textfont=sf]{subfig}
\usepackage{textcomp}
\usepackage{stfloats}
\usepackage{url}
\usepackage{verbatim}
\usepackage{graphicx}
\usepackage{cite}
\usepackage{indentfirst}
\usepackage{float}
\usepackage{subfigure}
\usepackage{multirow}
\usepackage{amsthm}
\usepackage{amsmath,amsthm}
\usepackage{cite}
\usepackage{amsmath,amssymb,amsfonts}
\usepackage{algorithmic}
\usepackage{graphicx}
\usepackage{textcomp}
\def\BibTeX{{\rm B\kern-.05em{\sc i\kern-.025em b}\kern-.08em
    T\kern-.1667em\lower.7ex\hbox{E}\kern-.125emX}}

\begin{document}
\title{The Communication GSC System with Energy Harvesting Nodes aided by Opportunistic Routing\\
\thanks{This work was supported by Natural Science Foundation of China (Project Number: U22B2003 and U2001208). (Corresponding author: Chen Dong).

Hanyu Liu, Lei Teng, Wannian An, Xiaoqi Qin, Chen Dong and Xiaodong Xu are with the School of Information and Communication Engineering Beijing University of Posts and Telecommunications. }
}
\author{\relax Hanyu Liu\IEEEauthorrefmark{2}, \relax Lei Teng\IEEEauthorrefmark{2}, \relax Wannian An\IEEEauthorrefmark{2}, \relax Xiaoqi Qin\IEEEauthorrefmark{2}, \relax Chen Dong\IEEEauthorrefmark{2}, \relax Xiaodong Xu\IEEEauthorrefmark{2}\\
\IEEEauthorblockA{\IEEEauthorrefmark{2}
State Key Laboratory of Networking and Switching Technology, Beijing, China\\
Beijing University of Posts and Telecommunications, Beijing, 100876, China\\ 
Email: hanyuliu@bupt.edu.cn, tenglei@bupt.edu.cn, anwannian2021@bupt.edu.cn, xiaoqiqin@bupt.edu.cn,\\
dongchen@bupt.edu.cn, xuxiaodong@bupt.edu.cn
}
}
\maketitle

\begin {abstract}
In this paper, a cooperative communication network based on energy-harvesting (EH) decode-and-forward (DF) relays is proposed. For relay nodes, there is harvest-storage-use (HSU) structure in this system. And energy can be obtained from the surrounding environment through energy buffering. In order to improve the performance of the communication system, the opportunistic routing algorithm and the generalized selection combining (GSC) algorithm are adopted in this communication system. In addition, from discrete-time continuous-state space Markov chain model (DCSMC), a theoretical expression of the energy limiting distribution stored in infinite buffers is derived. Through using the probability distribution and state transition matrix, the theoretical expressions of system outage probability, throughput and time cost of per packet are obtained. Through the simulation verification, the theoretical results are in good agreement with the simulation results.
\end{abstract}


\section{Introduction}

\IEEEPARstart{I}n recent years, the proposal of carbon peak and carbon neutrality makes sustainable development become the goal of social development \cite{1}. As a promising technology, energy-harvesting (EH) has raised researchers’ substantial concerns because of its capability of harvesting energy from the surrounding ambient energy source which consists of light energy, radio frequency (RF) energy, thermal energy \cite{2} and so on. Additionally, except for the advantages of energy storage, EH has longer battery lifetime in comparison of the battery which needs to be charged. For the reasons above, a number of articles casts light on the combination of EH  and takes insight into it.

The architecture of the EH nodes called HSU is taken into consideration. In literature \cite{3}, a simple communication system with EH using HSU architecture is taken into consideration. The EH node of the system harvests energy from a radio frequency signal broadcast by an access point in the downlink and uses the energy stored to transmit data in the uplink. And with help of EH and HSU, self-sustaining node (SSN) is taken into consideration in this paper. In literature \cite{4}, a system with relays is taken into consideration. And the relay is EH node. In this paper, considering the technology mentioned above, the relay nodes are EH nodes.

Opportunistic routing catches our attention because the systems mentioned above set the relays have same priority which may cause the efficiency of the system descend and the data could not find the best channel to transmit. Opportunistic routing is regarded as a solution to improve the performance of the wireless multi-top network. It has been widely studied in literature \cite{5}.

To select and combine the signals when the signal-to-noise ratio higher than a certain threshold, generalized selection combining (GSC) algorithm investigated in literature \cite{6} is exploited.  In recent study, the communication system using OR is introduced in literature \cite{7} and the communicaiton system using OR and maximal ratio combining (MRC) is introduced in \cite{8}. This paper aims to study OR-aided cooperative GSC communication networks with EH . Specifically, decode-and-forward (DF) relays are powered by harvested energy from the ambience using the infinite-size buffers with HSU architecture. Additionally, the DCSMC model is used. The main contributions of this work are as follows:
\begin{itemize}
\item[1)]
A cooperative network consists of energy-harvesting decode-and-forward relay nodes. The generalized selection combining (GSC) algorithm, OR algorithm are used. 
\end{itemize}
\begin{itemize}
\item[2)]
The state-transition-matrix based theoretical solutions and discrete-time continuous-state space Markov chain model are considered to get the limiting PDFs. 
\item[3)]
The closed-form expressions of outage probability and throughput are derived through analyzing the energy buffer storage and comparing signal-to-noise ratio with the threshold value.
\end{itemize}

\section{System Model,OR protocol}
\subsection{System Model consists of source node, relay nodes and destination node}

The network investigated in this paper consists of three sections: the source node S, the DF relay nodes, and a destination node D. All nodes mentioned above belong to half-duplex nodes. The nodes have different methods of energy supply. The source and destination nodes are provided energy by the power supply, while the R node is equipped with energy buffers using HSU architectures to apply energy-harvest technology to harvest ambient energy. Additionally, the quasi-static Rayleigh fading channel coefficients during the $i$-th time slot between S and D, S and R, R and D could be denoted by $\mathit{h_{SD}}\sim \mathcal{CN}$(0,$\mathit{d_{SD}^{-\alpha}}$), $\mathit{h_{SR}}\sim \mathcal{CN}$(0,$\mathit{d_{SR}^{-\alpha}}$), $\mathit{h_{RD}}\sim \mathcal{CN}$(0,$\mathit{d_{RD}^{-\alpha}}$). The expression like $A\sim \mathcal{CN}(0,\theta)$ means variable A follows the complex Gaussian distribution with mean 0 and variance $\theta$ and the $\mathit{d_{xy}}$ mentioned above denotes the distance between the node $x$ $\in$ $\lbrack$S,R$\rbrack$ and the node y $\in$ $\lbrack$R,D$\rbrack$, while the parameter $\alpha$ expresses the path-loss.

\begin{figure}[H]
\centering
\includegraphics[width=0.45 \textwidth]{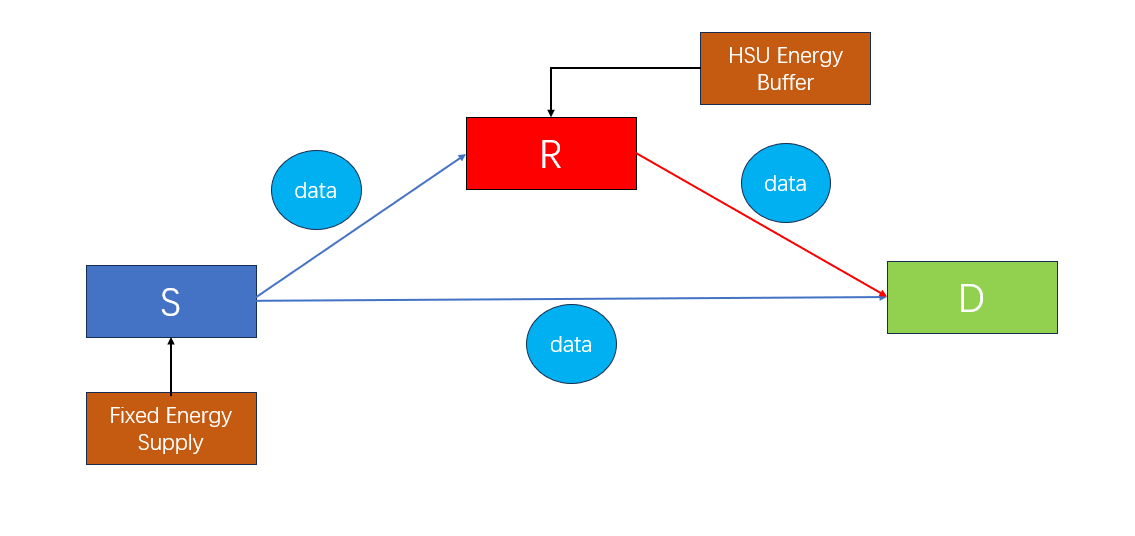}
\caption{System Model}
\label{Fig.5}
\end{figure}
\begin{figure}[H]
\centering
\includegraphics[width=0.45 \textwidth]{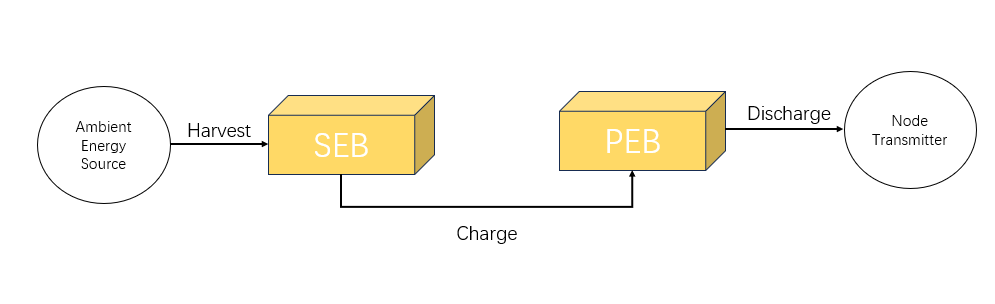}
\caption{HSU energy harvesting architecture}
\label{Fig.2}
\end{figure}

The HSU architecture equipped by the relay nodes has two energy buffers. In certain time slot, the energy harvested from ambiance would be stored in SEB at first. Then the SEB would transfer the energy mentioned above to the PEB at the end of the time slot to make the energy buffer equipped by relay node R charge and discharge simultaneously. For the reason that the energy harvested is much less than the actual energy buffer capacity, the capacity of the energy buffer is considered as a battery with infinite size in the theoretical analysis.

As the network woking process shown in Fig.1, the source noed S broadcasts unit-energy symbols $x_S$$(\mathit{i})$ to R and D at rate $R_0$ with the constant power $P_S$. And the received signals $\mathit{y}_{SR}$$(\mathit{i})$, $\mathit{y}_{SD}$$(\mathit{i})$ at R,D nodes in the  $\mathit{i}$-th time slot could be expressed by the following equations:
\begin{equation}
\mathit{y}_{SR}(\mathit{i})=\sqrt {Ps}\mathit{h_{SR}}(\mathit{i})x_S(\mathit{i})+\mathit{n_{SR}}(\mathit{i}) ,
\end{equation}
\begin{equation}
\mathit{y}_{SD}(\mathit{i})=\sqrt {Ps}\mathit{h_{SD}}(\mathit{i})x_S(\mathit{i})+\mathit{n_{SD}}(\mathit{i}),
\end{equation}
where, $\mathit{n_{SR}}$($\mathit{i}$) and $\mathit{n_{SD}}$($\mathit{i}$)$\sim\mathcal{CN}(0,\mathit{N_0})$ denote the received additive white Gaussian noese (AWGN) at R and D nodes respectively. Hence, the signal to noise ratios (SNRs) $\gamma_{SR}(\mathit{i})$ and $\gamma_{SD}(\mathit{i})$ at R and D in the $\mathit{i}$-th time slot are given as follow equations:
\begin{equation}
\gamma_{SR}(\mathit {i})=\frac{Ps \lvert \mathit{h_{SR}}^2 \rvert}{N_0},
\end{equation}
\begin{equation}
\gamma_{SD}(\mathit {i})=\frac{Ps \lvert \mathit{h_{SD}}^2 \rvert}{N_0},
\end{equation}

Since the use of a DF relay, R decodes the received signals and re-encodes them into unit energy symbols $x_{R}(\mathit{i})$, and then broadcasts $x_{R}(\mathit {i})$ to D with the constant power $M_{R}$. At last, the received signals $y_{RD}(\mathit {i})$ in the $\mathit{i}$-th time slot could be represented as follows:
\begin{equation}
\mathit{y}_{RD}(\mathit{i})=\sqrt {M_{R}}\mathit{h_{RD}}(\mathit{i})x_{R}(\mathit{i})+\mathit{n_{RD}}(\mathit{i}),
\end{equation}

where  $\mathit{n_{RD}}$($\mathit{i}$) $\sim\mathcal{CN}(0,\mathit{N_0})$ denotes the recieved AWGN at D respectively. Likely, the recieved SNR $\gamma_{RD}(\mathit{i})$ at D can be represented as follows:
\begin{equation}
\gamma_{RD}(\mathit {i})=\frac{M_{R} \lvert \mathit{h_{RD}}^2 \rvert}{N_0},
\end{equation}

Through the introduction of the channel assumption, the PDFs of SNRs of channel SR, SD and RD could be denoted as follows: 
\begin{equation}
f_{Node1Node2}(x) = W_{Node1Node2} e^{-W_{Node1Node2} x}.
\end{equation}

For example,the PDF of SNR of channel SR can be expressed as
$f_{SR}(x) = W_{SR} e^{-W_{SR} x}$, where, $W_{SD}=\frac{d_{SD}^{\alpha}N_0}{P_S}$. Similarly, $W_{SR}=\frac{d_{SR}^{\alpha}N_0}{P_S}$, $W_{RD}=\frac{d_{RD}^{\alpha}N_0}{M_{R}}$.
\begin{figure}[H]
\centering
\includegraphics[width=0.45 \textwidth]{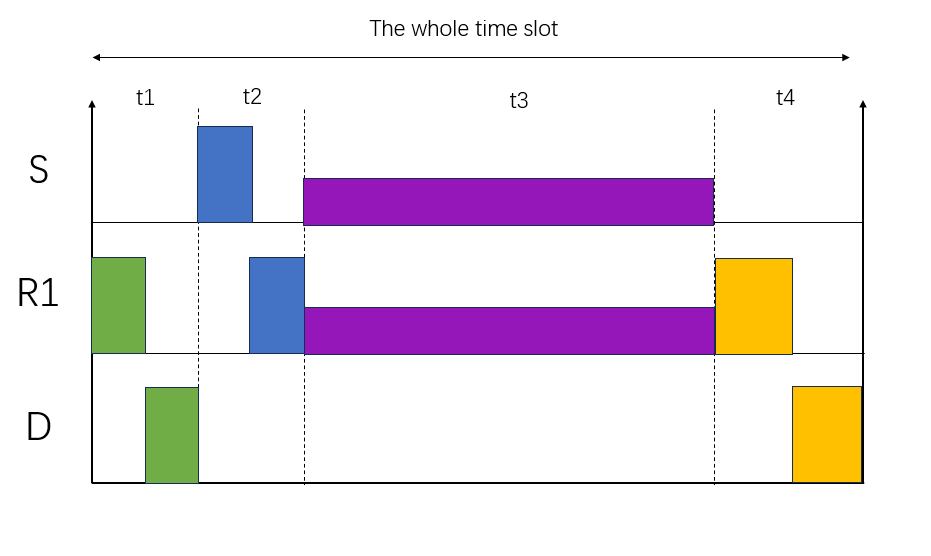}
\caption{The blocks in t1 denote pilot broadcasting; The  blocks in t2 denote signalling broadcasting; The blocks in t3 denote signalling broadcasting; The blocks in t4 denote packet broadcasting and the yellow clocks denote ACK/NACK broadcasting.  }
\label{Fig.2}
\end{figure}

A whole time slot consists of four sub time slots as shown in Fig.3. In t1, node D and R broadcast the pilot signals in order, making every node gain the CSI between itself and other nodes. In t2, every node estimates whether it is able to transmit the packet in view of its energy status, CSI, transmitter and receiver of the packet. The node will broadcast high-level signal when the node can transmit the packet; otherwise, it broadcasts low-level signal.  In t3, based on the transmission receiving above, nodes will determine whether transmit the packet by OR protocol introduced below this section. In this time slot, only one node can broadcast the packet. In t4, as receiver node, D and R will send a positive acknowledgment or a negative acknowledgment to show their reception status to correlation node. It is worth noting that ACK signals would be broadcast by the receiving node as the sign of successful packet reception only when the instantaneous link SNR at receiving node more than the threshold $\Gamma_{th}$.
\subsection{OR Protocol}
In this sub-section, the primary process based on OR protocol is investigated. In the network shown in Fig.1, S has the highest transmission priority and R has the lowest. The node which can receive the signals transmitted by the broadcast node is named the neighboring node. We assume that the neighboring nodes of the broadcast nodes of S are R and D, and the neighboring node of R is D. Determine the S(i) $\in \{\mathbf{s_{1}},\mathbf{s_{2}}\}$ shows the nodes which could transmit the packet in the current time slot, where $\mathbf{s_1}$=$\{S\}$, $\mathbf{s_2}$=$\{S,R\}$. In a timeslot, S will try to broadcast data to R and D. If D receives the data, S will try to send the next data. If D doesn't receive it and R receives it, S and R will transmit the data using GSC in the next timeslot. If R and D don't receive the data, S will wait until the SNR higher than the $\Gamma_{th}$. With the help of the CSI and energy-harvest technology, $\mathit{B_{1}(\mathit{i})}$ repesents the stored energy of R. If S and R have the data, S has higher transmission priority. S and R will use GSC to send data to D while R could transmit the data only enregy in buffer higher than the threshold. Determine the effective transmission and get S($\mathit{i}$+1) $\in \{\mathbf{s_{1}},\mathbf{s_{2}}\}$ in the next time slot.

\section{LIMITING DISTRIBUTION OF R ENERGY BUFFER}

In this section, the paper investigates the expressions of the limiting distributions of energy in the PEB of R and assumes it to be of infinite size. The harvest energy  input $X(i)$ in the $\mathit{i}$-th slot is assumed to be a variable obeying exponential distribution with probability density function (PDF) $\mathit{f}_{\mathit{X}}(x)$. And the mean of X($\mathit{i}$) is 1/$\lambda$. The PDF of X(i) could be denoted as follows:
\begin{equation}
\mathit{f}_{\mathit{X}}(x)=\lambda e^{-\lambda x},x>0.
\end{equation}
It is clear that  $\lambda_1$ denotes the parameter of the mean of X($\mathit{i}$) of R in this paper.

Since transmitting packets by broadcasting, D could select and combine the signals by GSC in different timeslot. With the use of GSC,  assume $\gamma_{\mathit{i}}$ represents the D received a signal with $\gamma_{\mathit{i}}$ SNR in previous timeslot. Through GSC, sum up the eligible $\gamma_{\mathit{i}}$ selected like $\gamma_{\mathit{overall}}=\sum\gamma_{\mathit{i}}$.
To show the derivation of the theoretical analysis more clearly and easier, letters are used in this process. The letters used in this paper are as follows:
\begin{equation}
\mathbf{A}\Leftrightarrow\gamma_{SD}(i)<\Gamma_{th},
\end{equation}
\begin{equation}
\mathbf{E}\Leftrightarrow \gamma_{overall}+\gamma_{SD}(i)<\Gamma_{th},
\end{equation}
\begin{equation}
\mathbf{G}\Leftrightarrow \gamma_{overall}+\gamma_{RD}(i)<\Gamma_{th},
\end{equation}
\begin{equation}
Pr\{\gamma_{SD}(i)<\Gamma_{th}\}=\mathbf{a},
\end{equation}
\begin{equation}
Pr\{\gamma_{overall}+\gamma_{SD}(i)<\Gamma_{th}\}=\mathbf{e},
\end{equation}
\begin{equation}
Pr\{\gamma_{overall}+\gamma_{RD}(i)<\Gamma_{th}\}=\mathbf{g}.
\end{equation}

The state of the transmitter nodes is important in the process of calculating the PDFs of the energy in PEB of R.  The $N \times N$ dimension state transition matrix (STM) T of two neighbouring time slots is used and each transmitter candidate state probability of occurrence which consists of $p_S=Pr\{S(i)=\mathbf{s_1}\}$, $p_{SR}=Pr\{S(i)=\mathbf{s_2}\}$. The elements of matrix T is denoted as follows:
\begin{equation}
T(i,j)=\mathbf{s_i}->\mathbf{s_j}\quad\quad\quad\quad(i,j\in[1,N]),
\end{equation}
in which the element $\mathbf{s_i}->\mathbf{s_j}(i,j\in[1,N])$ denotes the probability of the transmission from state i in the current time slot to state j in the next time slot. And the sum of the probability is equal to 1.  Through OR protocol shown above, we could get equations as follows:
\begin{equation}
	\begin{split}
		p_{\mathbf{s_1->s_1}} & = \mathrm{Pr}\{\gamma_{SD}(i)<\Gamma_{th}, \gamma_{SR}(i)<\Gamma_{th}\} \\
		& \quad + Pr\{\gamma_{SD}(i) \ge \Gamma_{th}\},
	\end{split}
\label{P_s}
\end{equation}
\begin{equation}
	\begin{split}
		p_{\mathbf{s_1->s_2}} & =\mathrm{Pr}\{\gamma_{SD}(i)<\Gamma_{th}, \gamma_{SR}(i)\ge\Gamma_{th}\},
	\end{split}
\end{equation}
\begin{equation}
	\begin{split}
		p_{\mathbf{s_2->s_1}} & = \mathrm{Pr}\{\overline{\mathbf{E}} \} + \mathrm{Pr}\{B_1(i) \ge M_{R}, \mathbf{E}, \overline{\mathbf{G}}\} ,
	\end{split}
\end{equation}
\begin{equation}
	\begin{split}
		p_{\mathbf{s_2->s_2}} & =1-p_{\mathbf{s_2->s_1}},
	\end{split}
\end{equation}
in which the parameter $PU1=\frac{1}{b_1\lambda_1 M_{R}} $. Parameters $b_1$ is given by the following equations. Additionally, the remaining transmissions not mentioned in T are equal to zero. 

Trough equations of the probability of transmissions, we could find out that the value of $\mathbf{e,g}$ is pivotal to get T. And through analysis of the equations of the probabilities, the probability which node D has received the values of SNRs at the state $S(i)$ is foundation to get the values of $\mathbf{e,g}$. And so the distribution $p_{overall}(j)$ is set as the probability distribution of $\gamma_{overall}$. With the use of GSC, the threshold of selection is set as $\mathbb{z}\Gamma_{th}$. For example, $p_{overall}(j)$ is the probability that D received the SNR of value which is denoted as follows:
\begin{equation}
\begin{split}
    \quad \quad&A_1(j) \leq \gamma_{overall} \leq A_2(j),\quad\quad\quad j\in[1,N-1],\\
    \quad \quad& 0 \leq \gamma_{overall} \leq \mathbb{z}\Gamma_{th},\quad\quad\quad\quad\quad j=0,
\end{split}             
\end{equation}
where, $A_1(j)=\mathbb{z}\Gamma_{th}+\frac{(j-1)\times(\Gamma_{th}-\mathbb{z}\Gamma_{th})}{N-1}$, $A_2(i)=\mathbb{z}\Gamma_{th}+\frac{j\times(\Gamma_{th}-\mathbb{z}\Gamma_{th})}{N-1}$, j $\in[1,N-1]$. And $A_1(0)=0$, $A_2(0)=\mathbb{z}\Gamma_{th}$.

In this section, STM method is used. STM $T_1$  represents $p_{overall}$. And the size of $T_1$ is also $N\times N$ dimension which is the same as T.  For example, the element $T_1(i,j)$ means that the transmission probability from ($A_1(i)\leq\gamma_{overall}<A_2(i)$) to ($A_1(j)\leq\gamma_{overall}\leq A_2(j)$), while 0 $\leq i , j \leq$ N-1 and $i , j \in Z$. Because the threshold value of GSC is set as $\mathbb{z}\Gamma_{th}$ in this paper and $\mathfrak{P}(0)$ means the probability of the SNRs D received less than threshold of GSC. So we could get $\mathfrak{P}(0)$=$\int_{0}^{\mathbb{z}\Gamma{th}}f_{SD}(x)dx$. So $T_1$ is as follows:
\vspace{-0.3cm}
\begin{equation}
	\begin{split}
		\textbf{T1}(i,j) &=(1-\int_{0}^{\Gamma_{th}-A_2(i)}f_{SD}(x)dx\\
		&\times(PU1\int_{0}^{\Gamma_{th}-A2(i)}f_{RD}(x)dx\\
		&\times \int_{A_1(j)}^{A_2(j)}f_{SD}(x)dx/\int_{0}^{\Gamma_{th}}f_{SD}(x)dx\\
		& \qquad \qquad \qquad \qquad \qquad \qquad \qquad \quad i \neq j
	\end{split}\label{T1}
\end{equation}
\begin{equation}
	\begin{split}
		\textbf{T1}(i,j) &=(1-\int_{0}^{\Gamma_{th}-A_2(i)}f_{SD}(x)dx\\
		&\times(PU1\int_{0}^{\Gamma_{th}-A_2(i)}f_{RD}(x)dx\\
		&\times \int_{A_1(j)}^{A_2(j)}f_{SD}(x)dx/\int_{0}^{\Gamma_{th}}f_{SD}(x)dx\\
		&+ \int_{0}^{\Gamma_{th}-A_2(i)}f_{SD}(x)dx\\
		&\times(PU1\int_{0}^{\Gamma_{th}-A_2(i)}f_{RD}(x)dx\\
		& \qquad \qquad \qquad \qquad \qquad \qquad \qquad \quad i = j
	\end{split}
\end{equation}

\begin{equation}
	\begin{split}
		p_{SD}(j)=\int_{A_1(j)}^{A_2(j)}f_{SD}(x)dx,\\
	\end{split}
\end{equation}
\begin{equation}
	\begin{split}
		p_{RD}(j)=\int_{A_1(j)}^{A_2(j)}f_{RD}(x)dx.\\
	\end{split}\label{T32}
\end{equation}

\addtolength{\topmargin}{0.04in}
  With the help of the equations and parameters analyzed above, the detailed values of $\mathbf{e,g}$ could be expressed as follows:
 \begin{equation}\label{eq.1}
	\begin{split}
		\mathbf{e}=\sum_{j=1}^{\mathcal{N}}\mathrm{conv}(p_{\gamma_{overall}},p_{SD}),
	\end{split}
\end{equation}
\begin{equation}
	\begin{split}
		\mathbf{g}=\sum_{j=1}^{\mathcal{N}}\mathrm{conv}(p_{\gamma_{overall}},p_{RD}),
	\end{split}
\end{equation}

while, conv(L,Z)denotes the discrete convolution such as $x(i)=conv(L,Z)=\sum\limits_{i=-\infty}\limits^{i=\infty}L(i)Z(j-i)$.
The process of obtaining parameters $p_s$, $p_{SR}$, $p_{\gamma_{overall}}(j)$, $\mathbf{e,g}$  are shown in $\mathbf{Algorithm 1}$.
\begin{algorithm}[ht!]
    \renewcommand{\algorithmicrequire}{\textbf{Input:}}
	\renewcommand{\algorithmicensure}{\textbf{Output:}}
	\caption{probability $p_s$, $p_{SR}$, $p_{\gamma_{overall}}(j)$, $\mathbf{e,g}$}
    \label{power}
    \begin{algorithmic}[1] 
        \REQUIRE  $W_{SD}$, $W_{RD}$, $W_{SR}$, $M_R$, $\Gamma_{th}$, $\lambda_1$; 
        
        \STATE Initializing n=0, k=0, n1=0, k1=0, $i$$\in$\{1,2\}  and 
$j=1$,\\ $\mathbf{p}(0)$=$[p_S(0), p_{SR}(0)]$. Every element of $\mathbf{p}(0)=1/2$

        \WHILE {1}
            \STATE $\textbf{p}_{\gamma_{overall}}(0)=[p_{\gamma_{overall}}(1),p_{\gamma_{overall}}(2)]$. Every element equal to $\frac{1}{2}$.
            \WHILE {1}
              \STATE   $\mathbf{p}_{\gamma overall}(n+1)$=$\mathbf{p}_{\gamma overall}(n)$$\mathbf{T_{1}}(n)$;
            \IF {$\mathbf{p_{\gamma overall}}(n)$ $\leq$ 0 $\lvert$ $\lvert$ $\mathbf{T_{j}(n)}\leq$ 0}
            \STATE $k = n-1;$,
                break,
            \ELSIF{$\lvert \mathbf{p_{overall}(i)}-\mathbf{p_{overall}(n+1)} \rvert \leq 10^{-7}$}
            \STATE $k=n$,
                break,
            \ELSE\STATE$n=n+1$,
            \ENDIF
             \ENDWHILE \\
	     Now calculate $\mathbf{p}$. The matrix $\mathbf{T}$ is given in (15).
	     \STATE   $\mathbf{p}(n1+1)$=$\mathbf{p}(n1)$$\mathbf{T}(n1)$, using the same method as calculating $\mathbf{p_{\gamma overall}}$ and then getting $\mathbf{p}(k1)$.
       \ENDWHILE
         \ENSURE $\mathbf{p}(k1)$, $\mathbf{p_{\gamma_{overall}}(k)}$, $\mathbf{e,g}$, $p_s$, $p_{SR}$ ;
    \end{algorithmic}
\end{algorithm}

Through the method mentioned in $\mathbf{Algorithm 1}$, the parameters above could be obtained. Now, the preparations are ready to give insight into the PDFs of energy in the PEB of R respectively.

	 By using the DCSMC and donating the energy level in the infinite-size energy buffer in the $i$-th signaling interval as $B_1(i)$, the PEB's buffer of R update equations are denoted as follows:
\begin{equation}\label{eq10}
	\begin{split}
		& B_1(i+1) = B_1(i)+X(i), \quad \quad {P}_{11}\\
		& B_1(i+1) = B_1(i)-M_{R}+X(i), \quad {P}_{12}
	\end{split}
\end{equation}
So $P_{11}$ and $P_{12}$ could be expressed as follows:
\begin{equation}\label{eq12}
	\begin{split}
		{P}_{12}:\, &\Big[\big(\mathbf{s_2} \big) \cap \big(B_1(i) \ge M_{R}\big)\cap \mathbf{E} \cap \overline{\mathbf{G}} \Big]\\
	\end{split}
\end{equation}
\begin{equation}\label{eq11}
	\begin{split}
		{P}_{11}:\, & \big(  \mathbf{s_1} \big)
		 \cup \Big[\big(\mathbf{s_2} \big) \cap \big(B_1(i) < M_{R}\big)\Big]\\
		& \cup \Big[\big(\mathbf{s_2} \big) \cap \big(B_1(i) \ge M_{R},\mathbf{E},\mathbf{G}\big)\Big],\\
        & \cup \Big[\big(\mathbf{s_2} \big) \cap \big(B_1(i) \ge M_{R},\overline{\mathbf{E}},\big)\Big].\\
	\end{split}
\end{equation}
As the method mentioned in last section, the stablity parameter of $B_1(i)$ could be expessed as follows:
\begin{equation}\label{eq22}
	\begin{split}
		\psi_{R}  & = \lambda_1 M_{R}\Big[p_{SR}\mathbf{e}((1-\mathbf{g})+\mathbf{g})\Big]
		= \lambda_1 M_{R}b_1 ,
	\end{split}
\end{equation}
where 
\begin{equation}\label{b_1}
	\begin{split}
		b_1  & = p_{SR}\mathbf{e}((1-\mathbf{g})+\mathbf{g})\\
	\end{split}
\end{equation}

	If $\psi_R > 1$, $B_1(i)$ in Eq. (\ref{eq10}) will have a stationary distribution. Furthermore, the limiting PDF of the energy buffer state at R can be expressed by
	\begin{equation}\label{eq23}
		g_1(x) = \begin{cases}
			\dfrac{1}{M_{R}}\left(1-e^{Q_1 x}\right), & 0 \leq x < M_{R} \\
			\dfrac{1}{M_{R}}\dfrac{-Q_1 e^{Q_1 x}}{\left(b_1\lambda_1+Q_1\right)}, & x \ge M_{R}.
		\end{cases}
	\end{equation}
\begin{equation}\label{Q_1}
	\begin{split}
		Q_1 = \frac{-W\left(-b_1\lambda_1 M_R e^{-b_1\lambda_1 M_{R}}\right)}{M_{R}}-b_1\lambda_1 < 0,
	\end{split}
\end{equation}
 satisfying $b_1 \lambda_1 e^{Q_1 M_{R}} = b_1 \lambda_1 + Q_1$.

\begin{proof}\renewcommand{\qedsymbol}{}
	The proof of the limiting PDF of R is given in Appendix A.
\end{proof}
\section{ANALYSIS OF OUTAGE PROBABILITY AND THROUGHPUT}
This section focuses on the performance of the system which consists of the analysis of failure probability, with the help of the analysis above and OR protocol,we could get:
\begin{equation}\label{SEC IV18}
	\begin{split}
		\mathrm{Pr}\{B_1(i) \ge M_{R}\}& =\dfrac{1}{b_1\lambda_1 M_{R}}=PU1.
	\end{split}
\end{equation}
And by the analysis of the failing transmission, the equation of failure probability(FP) is gotten
\begin{equation}\label{SEC IV19}
	\begin{split}
		FP & = p_S(1-e^{-W_{SD}\Gamma_{th}})\\
		&+p_{SR}\Big(\mathbf{e}(\mathbf{g}PU1+1-PU1)\Big)\\
	\end{split}
\end{equation}

\section{PERFORMANCE RESULTS}
In this paper, system investigated use the GSC technology, OR algorithm, while the nodes in this system belongs to EH nodes. Through these methods and technology, the system has lower failure probability, lower time cost and so on.

And the graphics of results are as follows:
\begin{figure}[H]
\centering
\includegraphics[width=0.45 \textwidth,angle=0]{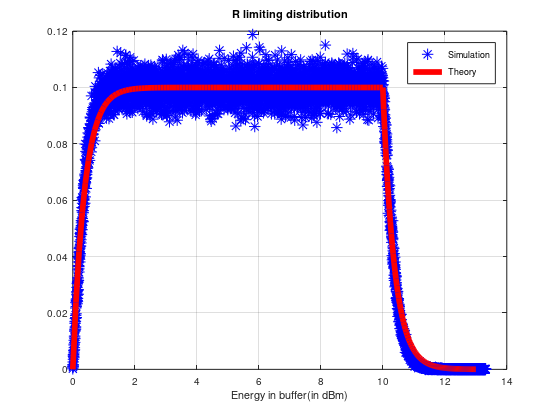}
\caption{The limiting distribution of R, with parameters $1/\lambda_1 = -11$ dB, $R_0=2$ bit/s/Hz, $M_R = 10$ mJoules, $\mathbb{z}=1/6$. S[0,0], R[45,20], D[100,0], $N_0$=-50dBm.}
\label{Fig.main5}
\end{figure}
In Fig.4, the PDF of the energy in buffer is shown. It shows that the simulaition is in line with the theory results. And when $x=M_{R}$, there is a turning point.

\begin{figure}[H]
\centering
\includegraphics[width=0.45 \textwidth,angle=0]{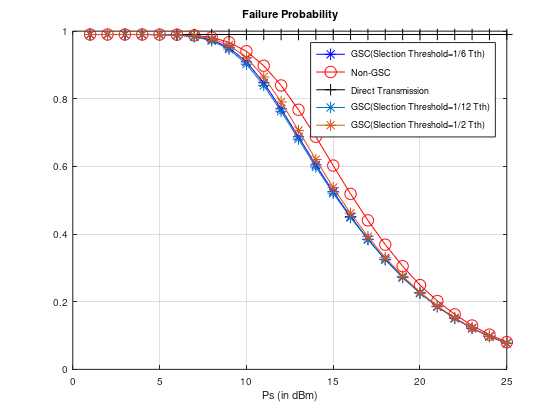}
\caption{The failure probability of each system, with parameters $1/\lambda_1 = -10$ dB $R_0=2$ bit/s/Hz, $M_R = 10$ mJoules, $\mathbb{z}=1/6, 1/2$ and $1/12$. S[0,0], R[45,20], D[100,0], $N_0$=-50dBm. }
\label{Fig.main5}
\end{figure}

In Fig.5, through the comparison with the system without GSC and the system using direct transmission, our system has lower fail probability while has larger throughput capacity.

\section{CONCLUSION}

This paper proposes the OR aided cooperative communication network with EH. GSC is used to improve the performance of the system. When $\mathbb{z}=0$, the system will change to a communication system with MRC. The OR protocol is proposed to select the packet transmission path based on the node transmission priority and the state of the energy in the energy buffer. In the future, enhance the performance of the system is our future research direction.
\appendices
\section{Proof of limiting distribution of R energy}
Through the process of investigating the probability ,  ${P}_{11}$ and ${P}_{12}$,  the CDF of $B_1(i+1)$ in Eq. (\ref{eq10}) can be expressed as follows:

\begin{equation}\label{APPENDIX B1}
	\begin{split}
		& \mathrm{Pr}\{B_1(i+1) \leq x\} \\
		&  = \mathrm{Pr}\{B_1(i)+X(i) \leq x, \mathbf{s_1}\} \\
		&  + \mathrm{Pr}\{B_1(i)+X(i) \leq x, \big(\mathbf{s_2} \big) , \big(B_1(i) < M_{R}\big)\}\\
		&  + \mathrm{Pr}\{B_1(i)+X(i) \leq x, \big(\mathbf{s_2} \big) , \big(B_1(i) \ge M_{R},\\
		&\qquad \mathbf{E},\mathbf{G}\big)\}+ \mathrm{Pr}\{B_1(i)-M_{R}+X(i) \leq x,\\
		&\qquad \big(\mathbf{s_2} \big) , \big(B_1(i) \ge M_{R}\big), \mathbf{E} , \overline{\mathbf{G}}\}\\
		&  + \mathrm{Pr}\{B_1(i)-M_{R}+X(i) \leq x,\big(\mathbf{s_2} \big) , \big(\big(B_1(i) \ge M_{R}\big), \mathbf{E} , \mathbf{G}\big)\\
	\end{split}
\end{equation}
When $i \to \infty$,  $\mathrm{Pr}\{B_1(i+1) \leq x\} = G^{i+1}_1(x) = G^{i}_1(x) = G_1(x)$ would have a stable state of energy buffer. Hence, Eq. (\ref{APPENDIX B1}) can be written as follows:
\begin{equation}\label{APPENDIX B5}
	\begin{split}
		G_{11}(x) & = \int_{\mu_1=0}^{x} F_X(x-\mu_1)g_1(\mu_1)\, d\mu_1 \\
		& \quad + b_1 \int_{\mu_1=M_{R}}^{x+M_{R}} F_X(x+M_{R}-\mu_1)g_1(\mu_1)\, d\mu_1, \\
		& \qquad \qquad \qquad \qquad \qquad \qquad \qquad 0\leq x < M_{R}
	\end{split}
\end{equation}
\begin{equation}\label{APPENDIX B6}
	\begin{split}
		G_{12}(x) & = \int_{\mu_1=0}^{M_{R}} F_X(x-\mu_1)g_1(\mu_1)\, d\mu_1 \\
		& \quad + a_1 \int_{\mu_1=M_{R}}^{x} F_X(x-\mu_1)g_1(\mu_1)\, d\mu_1 \\
		& \quad + b_1\int_{\mu_1=M_{R}}^{x+M_{R}} F_X(x+M_{R}-\mu_1)g_1(\mu_1)\, d\mu_1,\\
		& \qquad \qquad \qquad \qquad \qquad \qquad \qquad \quad x \ge M_{R}
	\end{split}
\end{equation}
where,
\begin{equation}\label{APPENDIX B9}
	\begin{split}
		a_1&=p_S+p_{SR}(\mathbf{e}\times\mathbf{g}+(1-\mathbf{e}))\\
	\end{split}
\end{equation}
\begin{equation}\label{APPENDIX B10}
	\begin{split}
		b_1  & =p_{SR}\times\mathbf{e}((1-\mathbf{g})+\mathbf{g})
	\end{split}
\end{equation}
Through the Eq.(\ref{APPENDIX B6}), the PDF $g_1(x)$ may be denoted as follows:
\begin{equation}\label{APPENDIX B8}
	\begin{split}
		g_{12}(x) & = \int_{\mu_1=0}^{M_{R}} f_X(x-\mu_1)g_{11}(\mu_1)\, d\mu_1 \\
		& \quad + a_1 \int_{\mu_1=M_{R}}^{x} f_X(x-\mu_1)g_{12}(\mu_1)\, d\mu_1\\ 
		& \quad+ b_1  \int_{\mu_1=M_{R}}^{x+M_{R}} f_X(x+M_{R}-\mu_1)g_{12}(\mu_1)\, d\mu_1,\\
		& \qquad \qquad \qquad \qquad \qquad \qquad \qquad \quad x \ge M_{R},
	\end{split}
\end{equation}
Similar to the process of calculating $g_{22}(x)$ in appendix B, let $g_{12}(x) = k_1 e^{Q_1 x}$. Substituting $g_{12}(x) = k_1 e^{Q_1 x}$ and $f_X(x) = \lambda_1 e^{-\lambda_1 x}$ into Eq. (\ref{APPENDIX B8}), we could gain
\begin{equation}\label{eq.134}
		\dfrac{b_1 \lambda_1 e^{Q_1 M_{R}} + a_1 \lambda_1}{\lambda_1 + Q_1} = 1
\end{equation}
\begin{equation}\label{eq.135}	
\begin{split}
\dfrac{k_1 e^{Q_1 M_{R}} \left[ b_1 + a_1 e^{\lambda_1 M_{R}}\right]}{\lambda_1 + Q_1}
=\int_{\mu_1=0}^{M_{R}}e^{\lambda_1 \mu_1}g_{11}(\mu_1)\,d\mu_1
\end{split}
\end{equation}
And the nonzero solution $Q_{1_1}$ of $Q_1$ in Eq. (\ref{eq.134}) can be obtained by simplifying Eq. (\ref{eq.134}) as
\begin{equation}\label{eq.136}
	b_1 \lambda_1 e^{Q_1 M_{R}} = \lambda_1 - a_1 \lambda_1 + Q_1= b_1 \lambda_1 + Q_1,
\end{equation}
Based on the Lambert W function, we obtain
\begin{equation}\label{APPENDIX B16}
\begin{split}
	Q_1 = \frac{-W\left(-b_1\lambda_1 M_{R} e^{-b_1\lambda_1 M_{R}}\right)}{M_{R}}-b_1\lambda_1,\\b_1\lambda_1 M_{R}>1.
\end{split}
\end{equation}

The derivatives of Eq. (\ref{APPENDIX B5}) about $x$ can be obtained
\begin{equation}\label{APPENDIX B17}
	\begin{split}
		g_{11}(x) & = b_1 \int_{\mu_1=M_{R}}^{x+M_{R}} f_X(x+M_{R}-\mu_1)g_{12}(\mu_1)\, d\mu_1\\
		& \quad + \int_{\mu_1=0}^{x} f_X(x-\mu_1)g_{11}(\mu_1)\, d\mu_1, 0 \leq x <M_{R}
	\end{split}
\end{equation}
Substituting $g_{12}(x)=k_1 e^{Q_1 x}$ and $f_X(x)=\lambda_1 e^{-\lambda_1 x}$ into Eq. (\ref{APPENDIX B17}), we get
\begin{equation}\label{APPENDIX B18}
	\begin{split}
		g_{11}(x) & = \lambda_1 \int_{\mu_1=0}^{x} e^{-\lambda_1 \left(x-\mu \right)}g_{11}(\mu_1)\, d\mu_1 \\
		& \quad + \dfrac{b_1 k_1 \lambda_1 e^{Q_1 M_{R}}}{\lambda_1 + Q_1} \left(e^{Q_1 x}-e^{-\lambda_1 x} \right), 0 \leq x <M_{R}
	\end{split}
\end{equation}
And the solution of $g_{11}(x)$ could be given as follows
\begin{equation}\label{APPENDIX B21}
	\begin{split}
		g_{11}(x) = \dfrac{b_1 k_1 \lambda_1 e^{Q_1 M_{R}}\left(e^{Q_1 x}-1 \right)}{Q_1}, 0 \leq x <M_{R}
	\end{split}
\end{equation}
According to the unit area condition on $g_1(x)$, we have
\begin{equation}\label{APPENDIX B22}
	\int_{x=0}^{\infty} g_1(x)\, dx = \int_{x=0}^{M_{R}} g_{11}(x)\, dx +\int_{x=M_{R}}^{\infty} g_{12}(x)\, dx = 1,
\end{equation}
Substituting $g_{11}(x) = \frac{b_1 k_1 \lambda_1 e^{Q_1 M_{R}}\left(e^{Q_1 x}-1 \right)}{Q_1}$ and $g_{12}(x) = k_1 e^{Q_1 x}$ into Eq. (\ref{APPENDIX B22}), we get
\begin{equation}\label{APPENDIX B23}
	\dfrac{b_1 k_1 \lambda_1 e^{Q_1 M_{R}}}{Q_1} \int_{x=0}^{M_{R}} \left(e^{Q_1 x}-1 \right)\, dx +  k_1 \int_{x=M_{R}}^{\infty} e^{Q_1 x}\, dx =1,
\end{equation}
Simplifying Eq. (\ref{APPENDIX B23}), we have
\begin{equation}\label{APPENDIX B24}
	\dfrac{b_1 k_1 \lambda_1 e^{Q_1 M_{R}}}{Q_1} \left[\dfrac{e^{Q_1 M_{R}}-1}{Q_1}-M_{R} \right] - \dfrac{k_1 e^{Q_1 M_{R}}}{Q_1}=1,
\end{equation}
Substituting Eq. (\ref{eq.136}) into Eq. (\ref{APPENDIX B24}), then simplifying Eq. (\ref{APPENDIX B24}), the value of $k_1$ can be obtained as follows
\begin{equation}\label{APPENDIX B25}
	k_1 = \dfrac{-Q_1}{M_{R} \left(b_1 \lambda_1 + Q_1\right)},
\end{equation}
Substituting Eq. (\ref{eq.136}) and Eq. (\ref{APPENDIX B25}) into Eq. (\ref{APPENDIX B21}), we arrive at
\begin{equation}\label{APPENDIX B26}
	g_{11}(x) =  \dfrac{1-e^{Q_1 x}}{M_{R}}.
\end{equation}
Substituting Eq. (\ref{APPENDIX B26}) into the right side of Eq. (\ref{eq.135}), we obtain
\begin{equation}\label{APPENDIX B27}
	\begin{split}
		\int_{\mu_1=0}^{M_{R}} e^{\lambda_1 \mu_1} g_{11}(\mu_1)\, d\mu_1
		& = \dfrac{1-e^{\left(\lambda_1+Q_1\right)M_{R}}}{\left(\lambda_1+Q_1\right)M_{R}} - \dfrac{1-e^{\lambda_1 M_{R}}}{\lambda_1 M_{R}},
	\end{split}
\end{equation}
The equation in Eq. (\ref{eq.136}) leads us to conclude $\lambda_1 M_{R} = \frac{\left(\lambda_1+Q_1\right)M_{R}}{b_1 e^{Q_1 M_{R}} + a_1}$. Substituting this conclusion in Eq. (\ref{APPENDIX B27}), we have
\begin{equation}\label{APPENDIX B29}
	\begin{split}
		\int_{\mu_1=0}^{M_{R}} e^{\lambda_1 \mu_1} g_{11}(\mu_1)\, d\mu_1
		& = \dfrac{k_1 e^{Q_1 M_{R}} \left(b_1+a_1 e^{\lambda_1 M_{R}}\right)}{\lambda_1+Q_1}.\\
	\end{split}
\end{equation}

Through the calculation above, the unique solutions of $g_{11}(x)$ and $g_{12}(x)$ are obtained.

\end{document}